\documentclass[aps,pra,reprint,superscriptaddress]{revtex4-1}
\usepackage{graphicx}
\usepackage{amsmath}
\usepackage{amssymb}
\usepackage{here}

\newcommand{\cconj}[1]{#1^{\ast}}
\newcommand{\hconj}[1]{#1^{\dag}}
\newcommand{\bvec}[1]{\boldsymbol{#1}}

\begin{document}

\title{Strongly spinor ferromagnetic Bose gases}
\author{Kohaku H. Z. So}
\affiliation{Department of Physics, University of Tokyo, 7-3-1 Hongo, Bunkyo-ku, Tokyo 113-0033, Japan}
%\thanks{}
\author{Masahito Ueda}
\affiliation{Department of Physics, University of Tokyo, 7-3-1 Hongo, Bunkyo-ku, Tokyo 113-0033, Japan}
\affiliation{RIKEN Center for Emergent Matter Science (CEMS), Wako, Saitama 351-0198, Japan}

\date{\today}

\begin{abstract}
	By studying the zero-temperature and nonzero-temperature phase diagrams of the ferromagnetic spin-1 Bose-Hubbard model under an external magnetic field,  
	we find that the competition between ferromagnetism and the quadratic Zeeman energy yields two superfluid phases, 
	which feature discontinuous first-order phase transitions between them for a strongly spinor Bose gas such as ${}^7$Li, 
	contrary to the corresponding continuum system. 
\end{abstract}

%\keywords{}

\maketitle

% If in two-column mode, this environment will change to single-column
% format so that long equations can be displayed. Use
%\begin{widetext}
%\end{widetext}

% \begin{figure}
% \end{figure}

% Surround figure environment with turnpage environment for landscape
% figure
% \begin{figure}
% \end{figure}

%\begin{acknowledgments}
%\end{acknowledgments}

% ****** End of file apstemplate.tex ******

%============================================================
\section{Introduction}
\label{sec:intro}

Ultracold atoms trapped in an optical lattice offer a unique research arena 
in which the physics of interacting quantum many-body systems can be explored with unprecedented controllability and precision \cite{BlochDalibardZwerger}.  
This is highlighted by the experimental realization of the superfluid to Mott-insulator phase transition \cite{GreinerEtal}  in the Bose-Hubbard model \cite{FisherFisher,JakschEtal,KrutitskyPhysRep}. 
Recently, ultracold atoms with spin degrees of freedom have attracted considerable interest \cite{KawaguchiUeda,StamperKurnUeda} 
due to the possibilities of simulating quantum magnetism and exploring the interplay between spatial and spin degrees of freedom \cite{LewensteinBook}. 
A minimal model for studying such physics is the spin-1 Bose-Hubbard model \cite{ImambekovLukinDemler},  
which describes spin-1 atoms in optical lattices. 
The model can be either ferromagnetic or antiferromagnetic, depending on the sign of the spin-dependent interaction. 
While the antiferromagnetic spin-1 Bose-Hubbard model has extensively been studied 
\cite{ImambekovLukinDemler,TsuchiyaKuriharaKimura,KimuraTsuchiyaKurihara,YamamotoEtal,JiangEtal} 
because of their distinct properties such as an even-odd parity effect in the superfluid to Mott-insulator phase transition, 
the ferromagnetic spinor Bose-Hubbard model has been studied less extensively; 
this is presumably because without an external magnetic field the model exhibits saturated ferromagnetism over the entire zero-temperature phase diagram \cite{KatsuraTasaki}, 
and properties of the system are similar to those of spinless bosons. 

However, in the presence of an external magnetic field, 
the competition between the quadratic Zeeman effect and the ferromagnetic interaction gives rise to rich phases \cite{KawaguchiUeda,StamperKurnUeda}. 
While the ferromagnetic spin-dependent interaction favors states with large magnetization, 
the positive quadratic Zeeman effect penalizes such spin states. 
In the continuum system, i.e., without optical lattices, this competition yields two superfluid phases with the continuous phase transition between them: 
the polar phase, in which all particles occupy the $\sigma = 0$ sublevel, 
and the broken-axisymmetry phase with nonvanishing transverse magnetization 
\cite{StengerEtal,MurataSaitoUeda,KawaguchiUeda,StamperKurnUeda}. 

To the best knowledge of the present authors, 
the phase diagrams of ferromagnetic spin-1 lattice bosons in the presence of the quadratic Zeeman effect remains largely unexplored. 
There is a study on the unit-filling Mott-insulator phase \cite{RodriguezEtal}; however,  
the interplay between spatial and spin degrees of freedom has yet to be studied.  
Experimental studies on the ferromagnetic spin-1 bosons have been carried out mainly with ${}^{87}$Rb, 
which has a very small spin-dependent interaction compared with spin-independent interaction. 
Quite recently, experimental efforts to realize spinor condensates of ${}^7$Li with a much larger spin-dependent interaction are underway \cite{VengalattorePrvCom}, 
which motivates us to explore the physics of strongly spinor Bose gases. 

In this paper, we study the zero-temperature and nonzero-temperature phase diagrams of 
the spin-1 ferromagnetic Bose-Hubbard model under an external magnetic field, 
and report discontinuous first-order phase transitions between two superfluid phases.  
It is known that for the ferromagnetic spin-1 Bose-Hubbard model without an external magnetic field, 
the ground state exhibits saturated ferromagnetism over the entire phase diagram, 
and phase transitions are continuous \cite{NatuPixleyDasSarma,deParnyEtalQMC}. 
In the presence of an external magnetic field, there are two superfluid phases:  
the polar superfluid phase and the broken-axisymmetry superfluid phase. 
We find that for the parameter region including the hyperfine spin-1 manifold of ${}^7$Li, 
the phase transitions between the broken-axisymmetry superfluid phase and the other phases are discontinuous for some part of the phase boundary, 
which makes a sharp contrast to the case without an external magnetic field. 
For ${}^7$Li, the discontinuous phase transition also occurs when the quadratic Zeeman energy is varied, 
contrary to the corresponding system without a lattice, in which the corresponding transition is predicted to be continuous 
on the basis of the mean-field theory for weak interaction \cite{KawaguchiUeda,StamperKurnUeda}. 

The rest of the paper is organized as follows. 
In Sec. II, we introduce the model we consider, i.e., the spin-1 Bose-Hubbard model with an external magnetic field, 
and describe the method of the decoupling approximation, which we employ to analyze our model. 
The obtained zero-temperature and nonzero-temperature phase diagrams are shown in Sec. III, 
where we see discontinuous phase transitions between the two superfluid phases. 
Based on these results, in Sec. IV, we compare our results with previous studies, and discuss their physical implications. 
In Sec. V, we conclude the paper. 
In the Appendix, we describe some details of numerical calculations. 

%============================================================
\section{The model and analyses}
\label{sec:manda}

%------------------------------------------------------------
\subsection{The model}
\label{subsec:manda_model}

We consider spin-1 bosons loaded in an optical lattice with an arbitrary geometry.
We denote by $a_{i,\sigma}$ the annihilation operator of a boson at lattice site $i$ and spin $\sigma \in \left\{ 1,0,-1 \right\}$. 
The particle-number operators are defined by 
$\hat{n}_{i,\sigma} := a_{i,\sigma}^{\dag} a_{i,\sigma}$ and $\hat{n}_{i} := \sum_{\sigma} \hat{n}_{i,\sigma}$, 
and the spin operators are defined by $S^{(\alpha)}_i := \sum_{\sigma,\sigma'} S^{(\alpha)}_{\sigma,\sigma'} a_{i,\sigma}^{\dag} a_{i,\sigma'}$ for $\alpha = x,y,z$, 
where $S^{(\alpha)}_{\sigma,\sigma'}$ denotes elements of spin-1 matrices $S^{(\alpha)}$ 
(for example, $S^{(z)}_{\sigma,\sigma'} = \sigma \delta_{\sigma,\sigma'}$. ).
We assume that an external magnetic field is applied along the quantization axis. 

Our system is described by the standard spin-1 Bose-Hubbard model \cite{ImambekovLukinDemler} whose grand Hamiltonian is given by 
\begin{eqnarray}
	\hat{K} &=& 
	- t \sum_{\langle i,j \rangle,  \sigma} \hconj{a}_{i,\sigma} a_{j,\sigma} 
	+ \frac{U_0}{2} \sum_{i} \hat{n}_i (\hat{n}_i - 1) \nonumber \\
	&{}& + \frac{U_1}{2} \sum_{i} \left( \boldsymbol{S}_{i}^{2} - 2 \hat{n}_i \right) 
	+ q \sum_{i ,\sigma} \sigma^2 \hat{n}_{i,\sigma} - \mu \sum_{i} \hat{n}_i, 
\end{eqnarray}
where $\mu$ is the chemical potential, and $\langle i,j \rangle$ means the sum over nearest neighbors. 
Here we distinguish $\langle i,j\rangle $ from $\langle j,i \rangle$ if $i \neq j $. 
We assume the positive hopping amplitude $t > 0$, the repulsive interaction $U_0>0$, and the ferromagnetic spin-dependent interaction $U_1 < 0$ 
as is the case for the $F=1$ ${}^{87} {\rm Rb}$ and the $F=1$ ${}^7 {\rm Li}$ (see table II of Ref. \cite{StamperKurnUeda}). 
The coefficient $q$ represents the quadratic Zeeman energy. 
The linear Zeeman energy can be ignored because the magnetization along the axis of the external magnetic field is conserved in this setting. 
Since we are interested in the competition between the quadratic Zeeman effect and the ferromagnetic interaction, 
we consider the case of $q > 0$, which holds true for various alkali atoms including ${}^7$Li and ${}^{87}$Rb with a static magnetic field.

%------------------------------------------------------------
\subsection{The decoupling approximation}
\label{subsec:manda_dc}

To analyze the system, we use the decoupling approximation \cite{SheshadriEtal,PaiSheshadriPandit}, 
which is equivalent to the Gutzwiller variational ansatz \cite{RokhsarKotliar,KrauthCaffarelBouchaud}. 
In this approximation, the hopping terms between different sites are decoupled as
\begin{eqnarray}
	a_{i,\sigma}^{\dag} a_{j,\sigma} 
	&=& (a_{i,\sigma}^{\dag}   -  \langle a_{i,\sigma}^{\dag} \rangle ) (  a_{j,\sigma} - \langle a_{j,\sigma} \rangle) \nonumber \\
	&{}& + \langle a_{i,\sigma}^{\dag} \rangle a_{j,\sigma}   + \langle a_{j,\sigma}\rangle a_{i,\sigma}^{\dag}   
		- \langle a_{i,\sigma}^{\dag} \rangle \langle a_{j,\sigma} \rangle  \\
	&\simeq &
\langle a_{i,\sigma}^{\dag} \rangle a_{j,\sigma}   + \langle a_{j,\sigma}\rangle a_{i,\sigma}^{\dag}   - \langle a_{i,\sigma}^{\dag} \rangle \langle a_{j,\sigma} \rangle, 
\end{eqnarray}
where $\langle \cdot \rangle$ represents the expectation value, 
and the term that is quadratic in the deviation from the expectation value is neglected in the second approximate equality. 
We assume that the state is spatially uniform so that $\phi_\sigma := \langle a_{i,\sigma} \rangle$ is independent of sites.
Then, different sites are decoupled and the problem reduces to a single-site problem with the grand Hamiltonian given by
\begin{eqnarray}
	\hat{K}(\bvec{\phi}) &=& 
	-zt \sum_{\sigma} 
		\left( \phi_\sigma \hconj{a}_{\sigma} + \cconj{\phi_\sigma} a_{\sigma} - |\phi_\sigma|^2 \right) 
		+ \frac{U_0}{2} \hat{n} (\hat{n} - 1) \nonumber \\
		&{}& + \frac{U_1}{2} \left( \bvec{S}^{2} - 2 \hat{n} \right) 
		+ q (\hat{n}_1 + \hat{n}_{-1})
		- \mu \hat{n}, 
		\label{eq:decoupledK}
\end{eqnarray}
where we omit the subscript $i$ for the lattice site,  
$z$ is the number of nearest neighbors, which is assumed to be independent of the lattice site, 
and $\bvec{\phi} := (\phi_1, \phi_0, \phi_{-1})$. 

The rationale behind the decoupling approximation is that, for both non-interacting ($U_0 = U_1 = 0$) and insulating ($t=0$) limits, 
the corresponding ground states can be written as products of single-site states, 
and hence it is natural to expect that the decoupled Hamiltonian can describe the superfluid to Mott-insulator phase transitions well. 
This method has been widely used to study both scalar \cite{SheshadriEtal, OostenEtal} and spin-1 Bose-Hubbard models 
\cite{TsuchiyaKuriharaKimura,JiangEtal,PaiSheshadriPandit,NatuPixleyDasSarma},  and
the comparison with quantum Monte Carlo simulations shows that 
the decoupling approximation can correctly capture the qualitative features of the phase transition for both scalar and spinor cases 
\cite{QMCscalar3D,QMCscalar2D,ClusterGutzwiller,deParnyEtalQMC}. 
We note that the approximation is known to be exact for bosons in the limit of infinite dimensions, i.e., the model with infinite range hopping \cite{RokhsarKotliar}. 

An approximate ground state or an equilibrium state can be obtained as a self-consistent solution to the single-site problem described above. 
For an approximate ground state, we seek for a single-site state vector $\psi$ satisfying the following conditions: 
(1) $\psi$ is the ground state of $\hat{K}(\bvec{\phi})$, and
(2) $\langle \psi , a_{\sigma} \psi \rangle = \phi_\sigma$ holds for all $\sigma$. 
For an approximate nonzero-temperature equilibrium state, we seek for a single-site density matrix $\rho$ satisfying the following conditions: 
(1) $\rho = e^{- \beta \hat{K}(\bvec{\phi}) } /  {\rm Tr} \left[ e^{- \beta \hat{K}(\bvec{\phi}) } \right]$, where $\beta$ is an inverse temperature, and
(2) $ {\rm Tr} \left( \rho a_{\sigma} \right) = \phi_\sigma$ holds for all $\sigma$. 

This procedure is carried out numerically by iteratively obtaining a state from $K(\bvec{\phi})$, 
calculating $\bvec{\phi}$ from the state, and plugging the obtained $\bvec{\phi}$ into $K$. 
To avoid being trapped by local solutions, we start the iteration from randomly chosen different initial values ($\bvec{\phi}$), 
and select the solution that has the lowest grand energy for the zero-temperature case or the lowest free energy 
$J(\bvec{\phi}) := -\frac{1}{\beta} \log {\rm Tr} \left[ e^{- \beta \hat{K}(\bvec{\phi}) } \right] $ for the nonzero-temperature case. 
In performing the numerical calculation, we truncate the Hilbert space by assuming the particle number per site is 
equal to or below a prescribed value $n_{\rm max}$. 
(The specific values of $n_{\rm max}$ will be shown in Sec. \ref{sec:results}. )
Other details of the numerical calculation are shown in the appendix.

The self-consistent conditions can be paraphrased in terms of the grand-energy function 
$E(\bvec{\phi}) := \min_{\psi} \langle \psi , \hat{K}(\bvec{\phi})  \psi \rangle $ for the zero-temperature case,
 or the free energy $J(\bvec{\phi})$ for the nonzero-temperature case. 
The superfluid order parameters $\boldsymbol{\phi}$ that satisfy the self-consistent condition are equivalent to stationary points of $E$ for the zero-temperature case, 
or $J$ for the nonzero-temperature case
(see Sec. 3.1.1 of \cite{WagnerThesis} for detail). 
Thus, $\bvec{\phi}$ corresponding to the approximate ground (or the thermal equilibrium) state 
can be identified with a minimum of $E(\bvec{\phi})$ (or that of $J(\bvec{\phi})$). 
This property will be used to clarify the discontinuous phase transitions.

%============================================================
\section{Phase diagrams}
\label{sec:results}

%------------------------------------------------------------
\subsection{Zero-temperature phase diagrams}
\label{subsec:results_gs}

In this section, we discuss zero-temperature phase diagrams. 
In calculating results throughout this subsection, the maximum number of particles per site, $n_{\rm max}$, 
is taken to be $n_{\rm max} = 7$, which is confirmed to be sufficient for numerical convergence in the parameter regime studied here. 

\begin{figure}
	\includegraphics[width = 9cm]{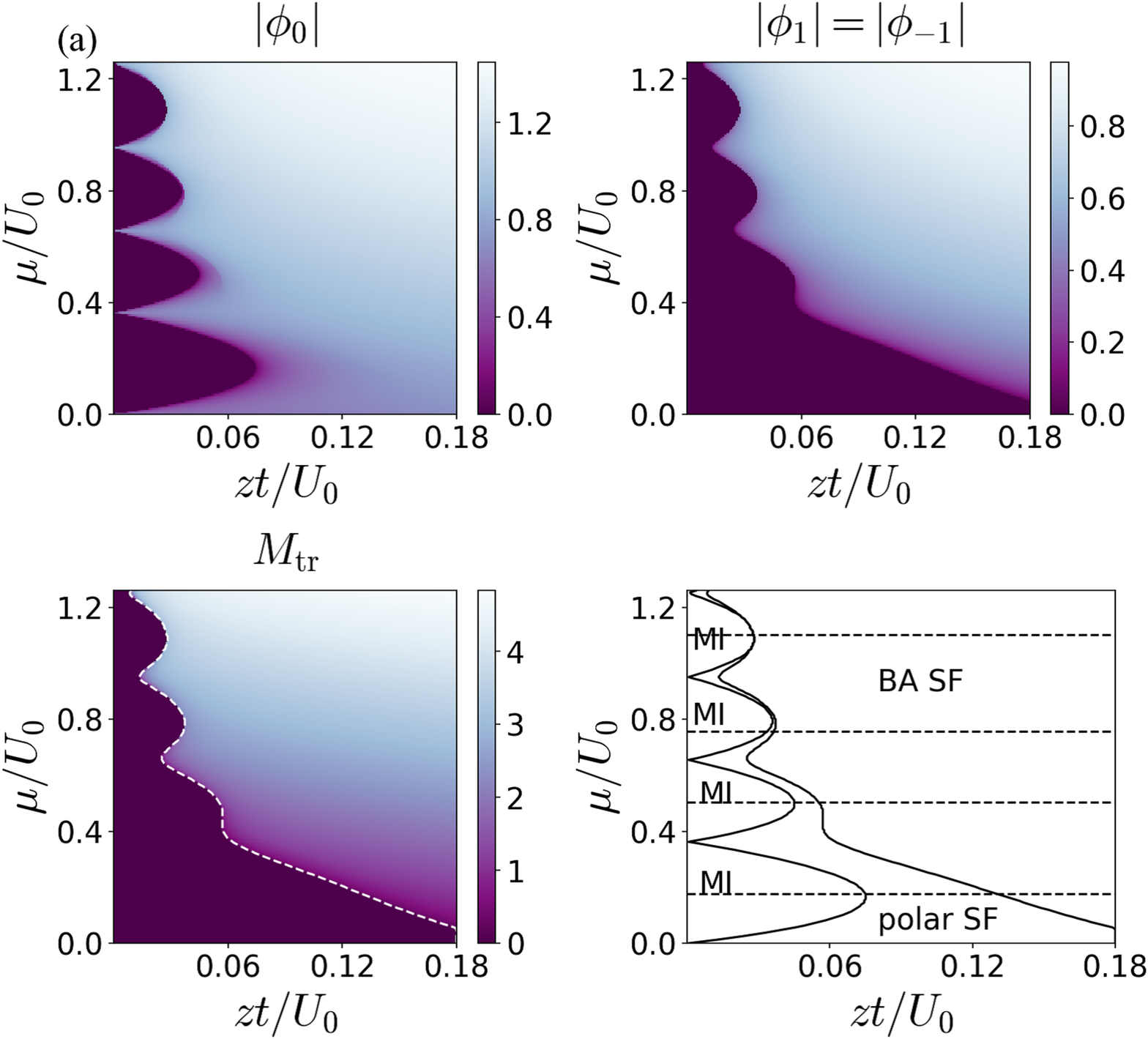}
	\includegraphics[width = 9cm]{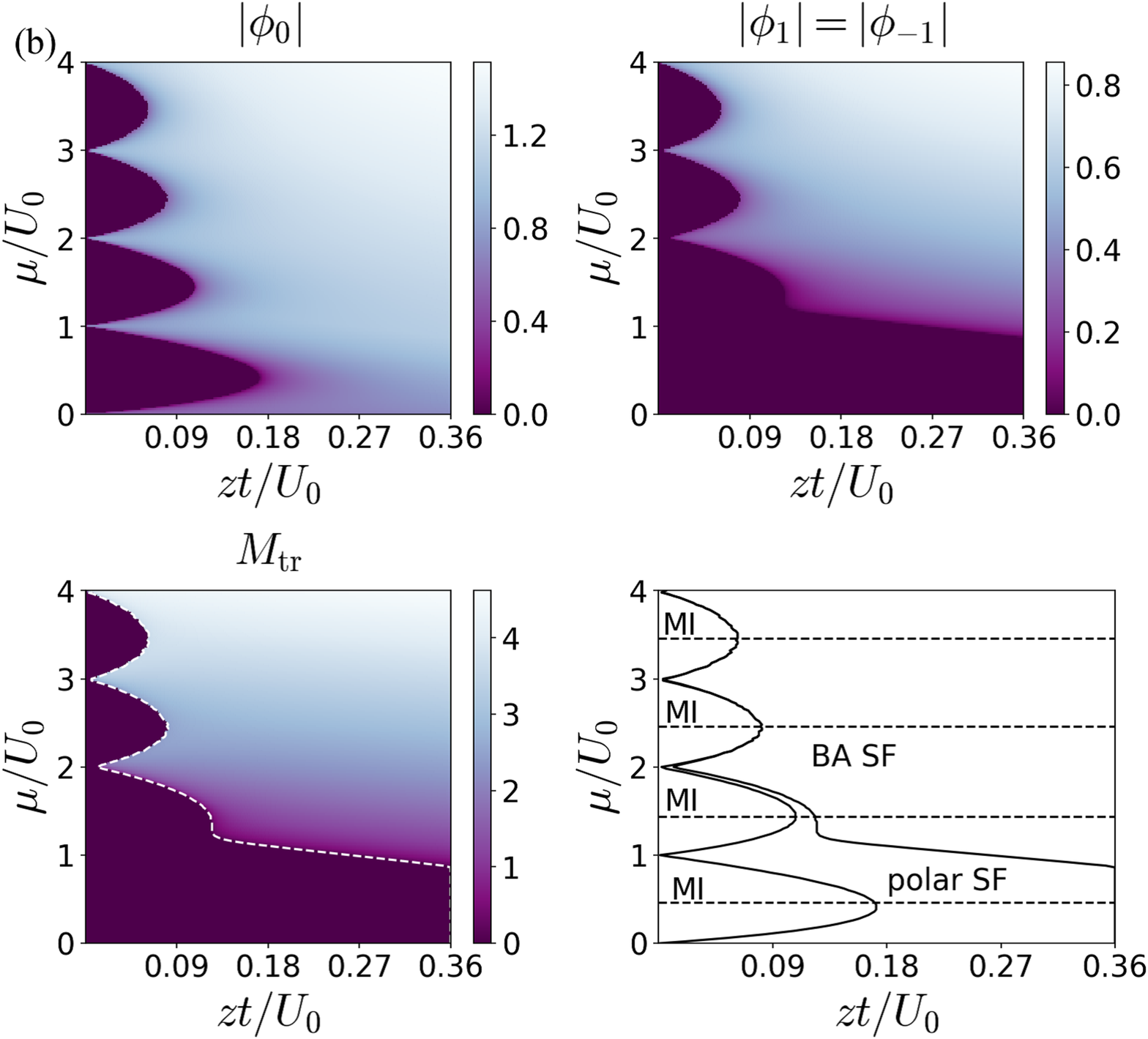}
	\caption{
		Zero-temperature phase diagrams for (a) $(U_1/U_0, q/U_0)  = (-0.7, 0.1)$ (${}^7$Li) and (b) $(-0.005, 0.0085)$ (${}^{87}$Rb). 
		For both (a) and (b), the upper left and right panels show $|\phi_0|$ and $|\phi_1|$, respectively.
		($|\phi_{-1}|$ is the same with $|\phi_1|$ within numerical errors.), 
		and the lower left and right panels show the transverse magnetization 
		$M_{\rm tr}$, 
		and the entire phase diagram, respectively. 
		In the lower right panel, dashed lines indicate the slices taken in Fig. \ref{fig:slice}, 
		``MI'' stands for Mott-insulator, 
		``SF'' stands for superfluid, and 
		``BA'' stands for broken-axisymmetry. 
		In the panels of the transverse magnetization, the white dashed curve shows the boundary calculated from the data of $\phi_1$, 
		which coincides with the boundary across which $M_{\rm tr}$ rises from zero. 
		Contrary to the case without the quadratic Zeeman effect, there are two (polar and broken-axisymmetry) superfluid phases, 
		and the lower-left plot shows that the transverse magnetization arises in the broken-axisymmetry superfluid phase. 
		The magnetization parallel to an applied magnetic field $\langle S^{(z)} \rangle$  is always zero within numerical accuracy. 
		}
	\label{fig:SFOPs}
\end{figure}

Figure \ref{fig:SFOPs} shows the obtained phase diagrams for $(U_1/U_0, q/U_0)  = (-0.7, 0.1)$ and $(-0.005, 0.0085)$, 
where the former set of parameters are for  ${}^7$Li \cite{VengalattorePrvCom}, and the latter for ${}^{87}$Rb (see table 2 of \cite{KawaguchiUeda}). 
No obtained ground states exhibit magnetization in the direction of the external magnetic field within numerical accuracy. 
In both cases, the phase diagram is significantly altered compared with the $q=0$ case: there appear two superfluid phases. 
In the superfluid phase near to the Mott lobes, only the $\sigma=0$ component shows superfluidity, which we identify as the polar superfluid phase.
The other phase has three superfluid components, and possesses a non-zero transverse magnetization 
$M_{\rm tr} := \sqrt{ \left\langle S^{(x)} \right\rangle^2 + \left\langle S^{(y)} \right\rangle^2 }$; 
hence we identify it as the broken-axisymmetry superfluid phase. 
With increasing $q$, the region of the polar superfluid phase expands, 
which is natural since positive $q$ penalizes the $\sigma = \pm 1$ components. 
Qualitatively the same behavior is obtained for different values of $U_1/U_0$. 

\begin{figure*}
	\begin{tabular}{c}
		\includegraphics[width=13cm]{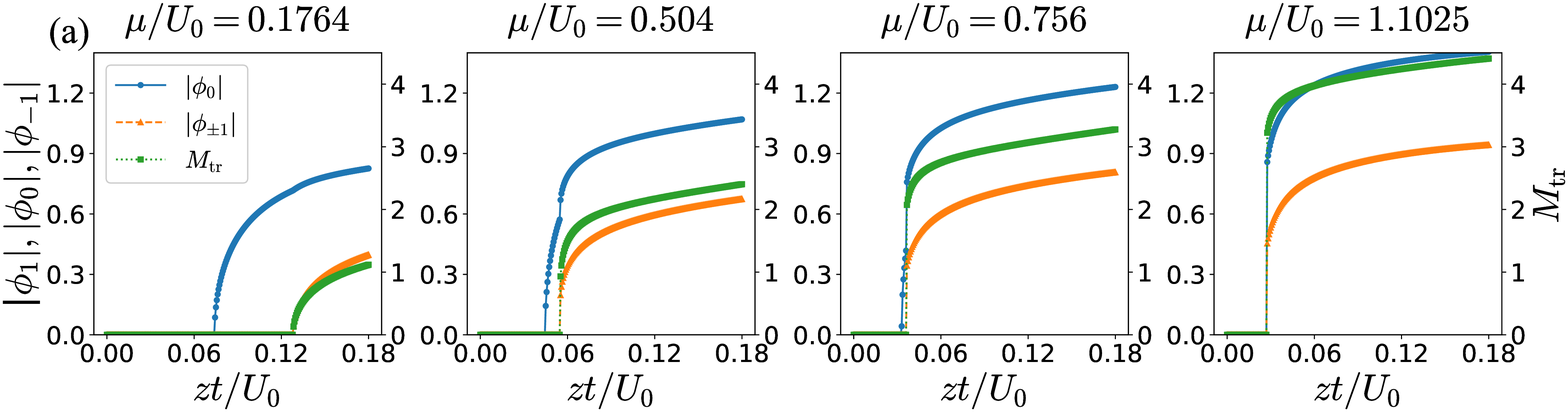}  \\
		\includegraphics[width=13cm]{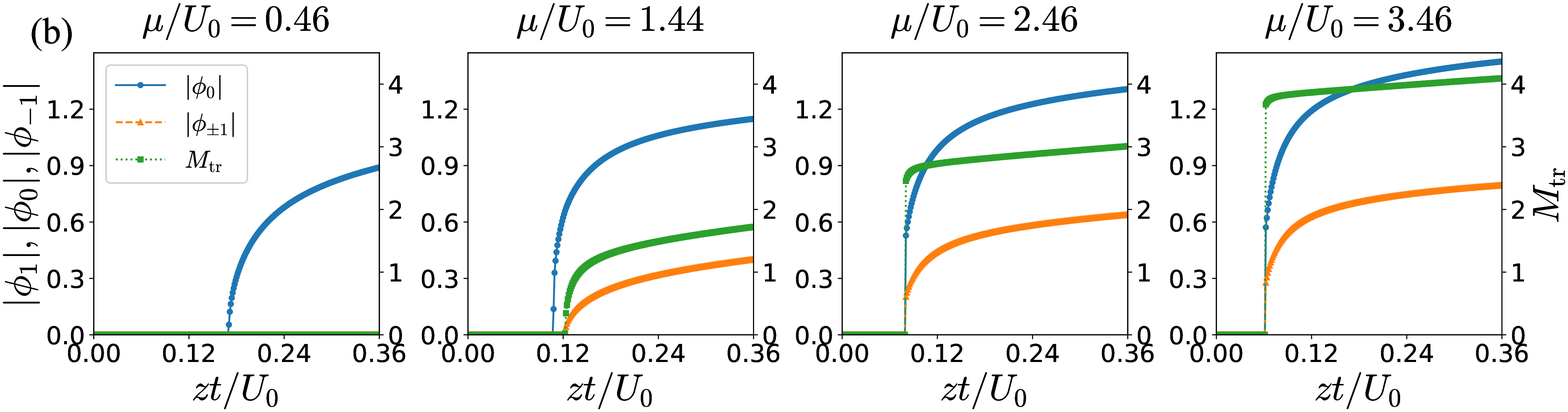} 
	\end{tabular}
	\begin{tabular}{c}
		\includegraphics[width=4.5cm]{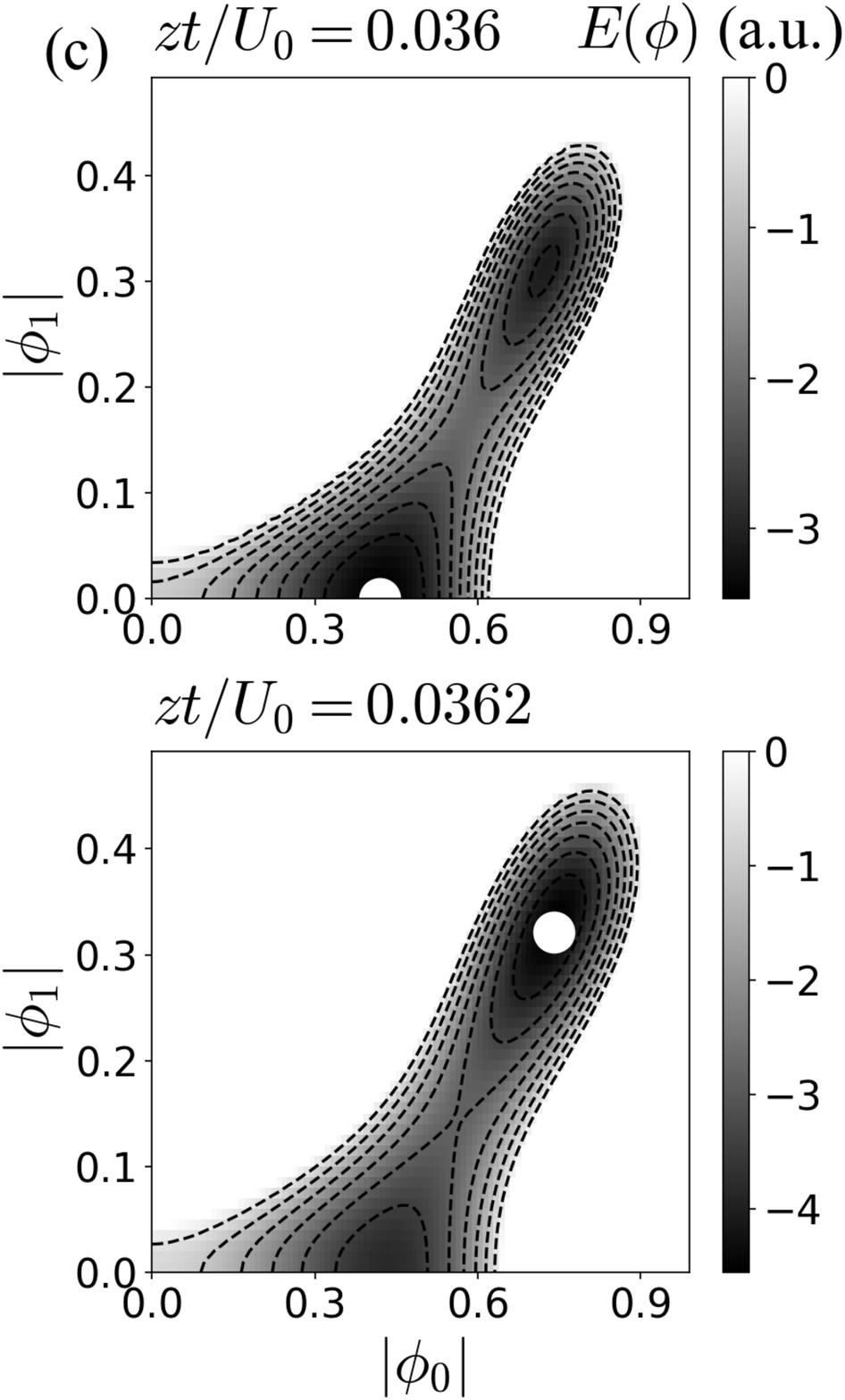}
	\end{tabular}
	\caption{(a) and (b) : 
		Dependences of order parameters on the hopping amplitude $t$ for fixed chemical potential $\mu$. 
		The left scale shows the magnitude of superfluid order parameters $|\phi_{\sigma}|$, 
		and the right one shows the transverse magnetization $M_{\rm tr}$.  
		The parameters used are (a) $U_1/U_0 = -0.7$, $q/U_0 = 0.1$, and (b) $U_1/U_0 = -0.005$, $q/U_0 = 0.0085$. 
		For phase transitions between the polar and broken-axisymmetry superfluid phases, 
		the panels in (a)  show discontinuous phase transitions, while panels in (b) show continuous phase transitions. 
		Discontinuous phase transitions between the broken-axissymmetry superfluid phase and the Mott-insulator phase should be interpreted with caution (see the text for details). 
		(c) : Dependence of the grand-energy function $E(\bvec{\phi}) := \min_{\psi} \langle \psi , \hat{K}(\bvec{\phi})  \psi \rangle $ 
		on $|\phi_0|$ and $|\phi_1|$, 
		where, by the assumption of the inversion symmetry $|\phi_1| = |\phi_{-1}|$, 
		$\bvec{\phi}$ can be written as $(|\phi_1|, e^{i\theta}, |\phi_0|, |\phi_1| e^{i\theta})$ and the minimum with respect to $\theta$ has already been taken numerically. 
		White circles indicate global minima. 
		For clarity, values above a given threshold value are not plotted, and the grayscale bars are shown in an arbitrary unit. 
		It can be seen that there are two local minima, indicating the metastability and the ensuing discontinuous phase transition. 
		The parameters used are $U_1/U_0 = -0.7, q/U_0 = 0.1$ and $\mu/U_0 = 0.756$. 
		}
	\label{fig:slice}
\end{figure*}

Although the obtained $t-\mu$ phase diagrams for ${}^7$Li and ${}^{87}$Rb apparently look similar, 
a closer look reveals a few important differences. 
In Fig. \ref{fig:slice} (a) and (b), we show the $t$ dependence of the order parameters $\phi_{\sigma}$ and $M_{\rm tr}$, 
where the left and right scales refer to $\phi_{\sigma}$ and $M_{\rm tr}$, respectively. 
For ${}^7$Li (panel (a)), 
transitions between the broken-axisymmetry superfluid phase and the polar superfluid phase are discontinuous for some part of the boundary 
(see plots for $\mu/U_0 = 0.504, 0.756$). 
The transverse magnetization also jumps at the transition point. 
For ${}^{87}$Rb, we find no discontinuous phase transitions between two superfluid phases in the parameter region we have explored. 
Although we have discontinuous phase transition between the broken-axisymmetry superfluid phase and the Mott-insulator phase for both cases, 
this result should be interpreted with caution (See subsection \ref{subsec:disc_MI}.).

To confirm the discontinuous phase transitions and the accompanying metastability for ${}^7$Li, we calculate 
$E(\bvec{\phi}) := \min_{\psi} \langle \psi , \hat{K}(\bvec{\phi})  \psi \rangle $ 
as a function of $\bvec{\phi}$. 
By assuming the inversion symmetry $|\phi_1| = |\phi_{-1}|$ 
and using a global phase multiplication and a rotation around the $z$-axis, we can restrict ourselves to the case with 
$\bvec{\phi} = (|\phi_1| e^{i\theta}, |\phi_0|, |\phi_1| e^{i\theta})$, where $\theta \in \mathbb{R}$. 
Plotted in Fig. \ref{fig:slice} (c) is $E(\bvec{\phi})$ as a function of $|\phi_1|$ and $|\phi_0|$, where the minimum with respect to $\theta$ has already been taken. 
We can see two local minima, which implies the metastability and the ensuing discontinuous phase transition. 

\begin{figure}
	\includegraphics[width=8cm]{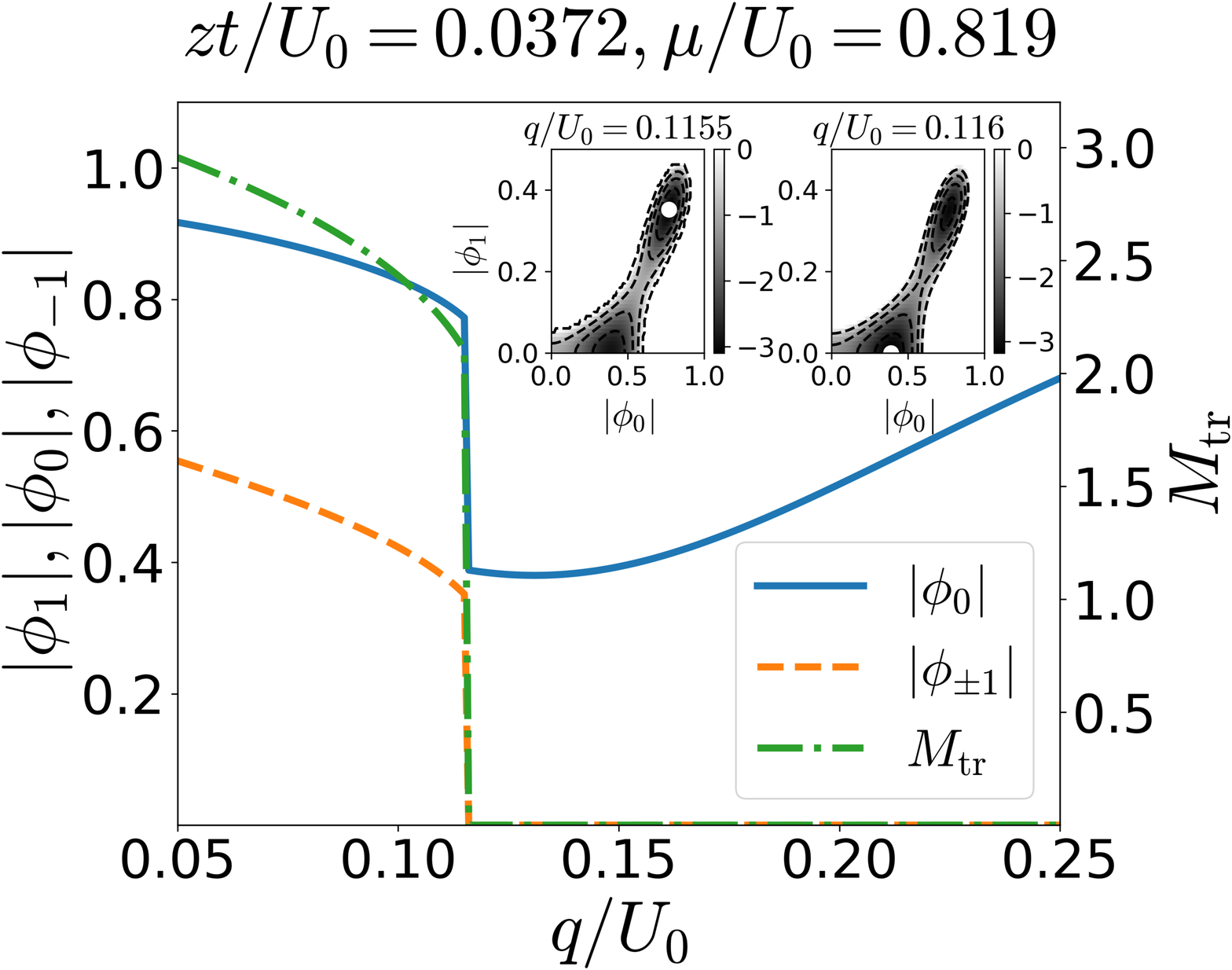}
	\caption{Superfluid order parameters (left scale) and transverse magnetization (right scale) as functions of $q/U_0$ 
		for $U_1/U_0 = -0.7, zt/U_0 = 0.0372$, and $\mu/U_0 = 0.819$. 
		The discontinuous phase transition occurs at $q/U_0 \simeq 0.116$. 
		The inset shows the grand-energy function $E(\bvec{\phi})$ as a function of $\phi_0$ and $|\phi_1|$, 
		which signifies the metastability (see the caption of figure \ref{fig:slice} for detail). }
	\label{fig:Qdependence}
\end{figure}

Next, we consider the transition due to a change in $q$. 
Figure \ref{fig:Qdependence} shows the $q$-dependence of the order parameter $\phi_{\sigma}$ and that of the transverse magnetization, 
which shows the discontinuous transition between the broken-axisymmetry superfluid phase and the polar superfluid phase. 

So far, we have seen that within our approximation scheme of the decoupling approximation 
the spin-1 Bose-Hubbard model with the quadratic Zeeman energy exhibits discontinuous phase transitions 
between the broken-axisymmetry superfluid phase and the other (polar and Mott-insulator) phases. 
Based on this observation, in Fig. \ref{fig:boundary}, we show the phase boundaries obtained by the binary search method described in the appendix. 
It can be seen that the discontinuous phase transitions occur 
only when the transition points are close to the Mott-insulator phase. 
\begin{figure}
	\includegraphics[width=8cm]{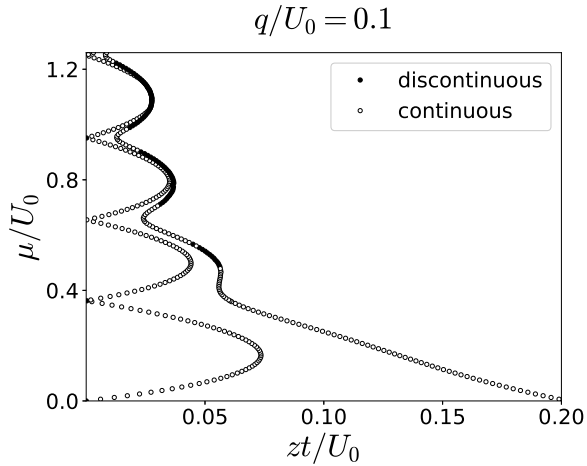}
	\caption{
	The $t-\mu$ zero-temperature phase diagram for $(U_1/U_0, q/U_0) = (-0.7, 0.1)$, 
	where the phase boundaries are determined by the binary search algorithm described in the appendix.  
	White circles show the boundary across which continuous phase transitions occur, 
	while black circles show the boundary across which discontinuous phase transitions occur. 
	It can be seen that discontinuous phase transitions between two superfluid phases occur when the boundary is close to the Mott lobes.}
	\label{fig:boundary}
\end{figure}

%------------------------------------------------------------
\subsection{Nonzero-temperature phase diagram}
\label{subsec:results_nzt}

In this subsection, we show the results for the nonzero-temperature case, and compare them with those for the zero-temperature case. 
For sufficiently low temperature, we see that the system exhibits qualitatively the same behavior as the zero-temperature, 
i.e., the two superfluid phases and discontinuous phase transitions between them. 
However, we also see that with the increase of temperature, the region of discontinuous phase transitions shrinks, and finally disappears. 
The implication of this section on experiments will be discussed in Sec. IV, C. 
Throughout this section, we consider smaller particle-number densities compared with the previous section to decrease the numerical cost. 
For that aim, we take $\mu$ to be small compared with that used in the previous section, and set $n_{\rm max} = 5$ and $(U_1/U_0, q/U_0) = (-0.7, 0.02)$. 
The choice $n_{\rm max} = 5$ is confirmed to be sufficient. 

\begin{figure}
	\includegraphics[width = 8cm]{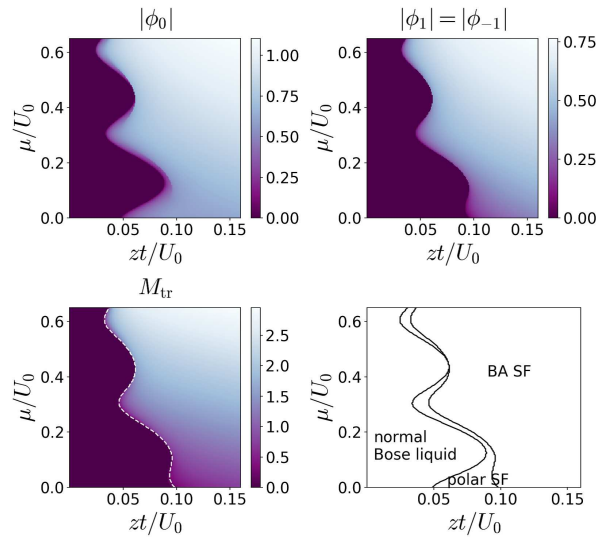}
	\caption{
		Nonzero-temperature equilibrium phase diagrams for $(U_1/U_0, q/U_0, k_B T/ U_0) = (-0.7, 0.02, 0.02)$ (${}^7$Li). 
		Upper left: $|\phi_0|$, upper right: $|\phi_1|$, 
		lower left: transverse magnetization $M_{\rm tr}$, 
		where the white dashed curve shows the boundary obtained from the data of $\phi_1$, 
		which coincides with the boundary across which $M_{\rm tr}$ arises from zero,  
		and lower right: the entire phase diagram, 
		where SF stands for superfluid. 
		The magnetization parallel to the applied magnetic field is always zero (within numerical accuracy). 
		}
	\label{fig:results_SFOPs_T>0}
\end{figure}

Figure \ref{fig:results_SFOPs_T>0} shows the obtained observables and the phase diagram in the $t-\mu$ plane, where the temperature is $k_B T/U_0 = 0.02$.  
Similarly to the zero-temperature case, the system shows two superfluid phases, i.e., the polar superfluid phase and the broken-axisymmetry phase. 

\begin{figure}
	\centering
	\includegraphics[width=8cm]{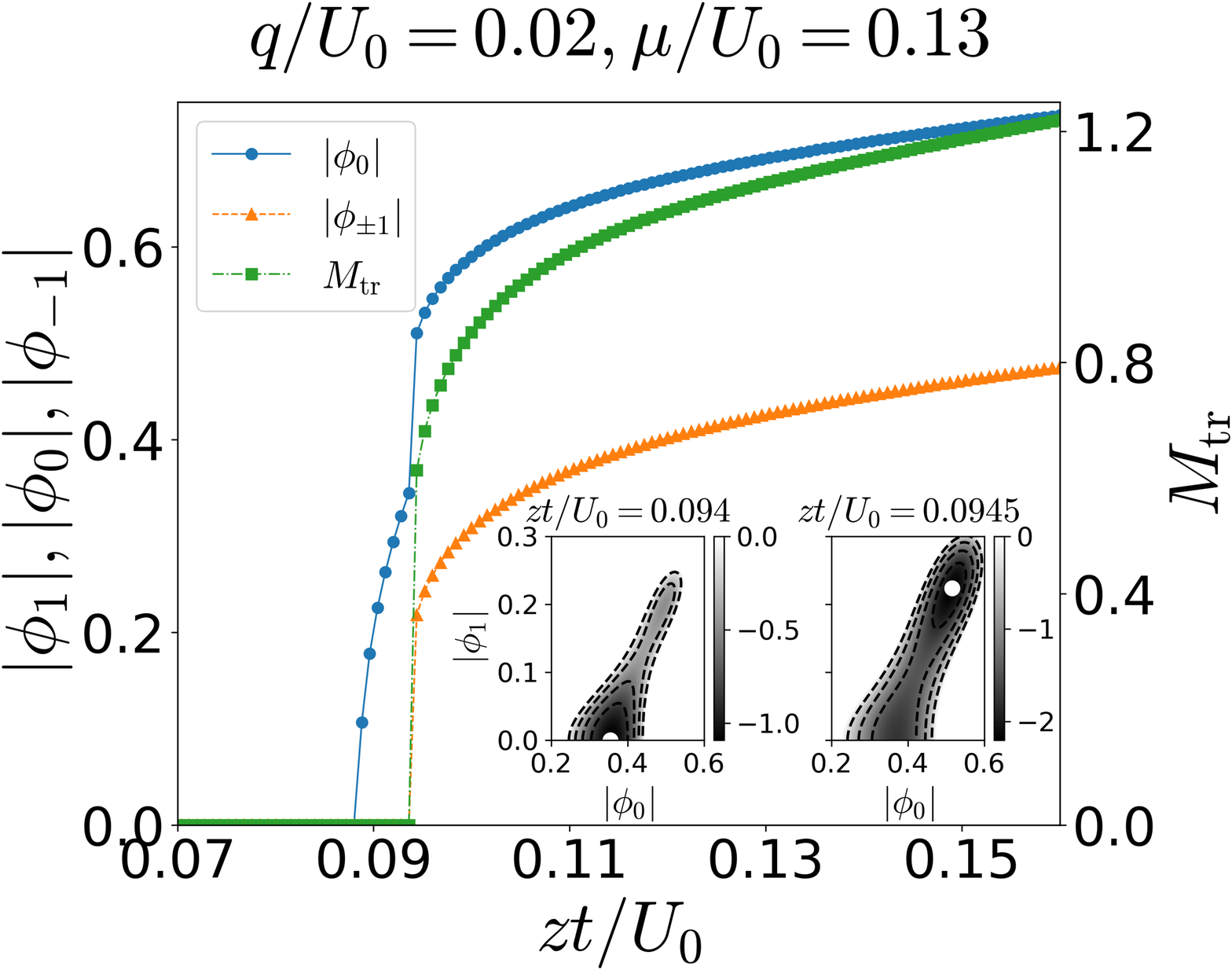}
	\caption{
		Superfluid order parameters (left scale) and transverse magnetization (right scale) as functions of $zt/U_0$ at nonzero temperature.  
		A discontinuous phase transition occurs at $zt/U_0 \simeq 0.094$. 
		The inset shows the thermodynamic potential $J(\bvec{\phi})$, which signifies the metastability (see the caption of Fig. \ref{fig:slice} for detail).
		Parameters are $U_1/U_0 = -0.7, q/U_0 = 0.02, \mu/U_0 = 0.13$, and $k_B T / U_0 = 0.02$. 
	}	
	\label{fig:tdependence_T>0}
\end{figure}

\begin{figure}
	\centering
	\includegraphics[width=8cm]{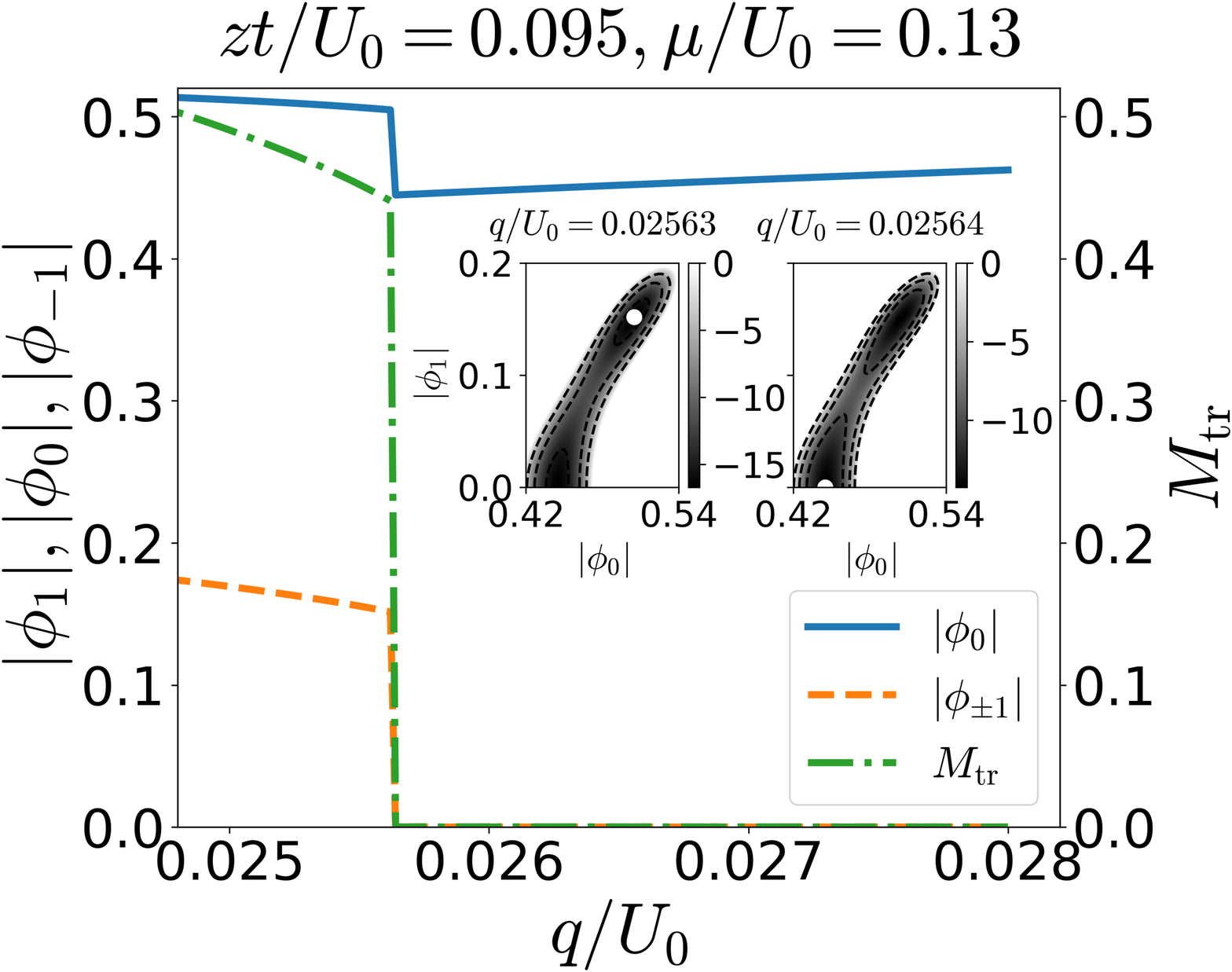}
	\caption{Superfluid order parameters (left scale) and transverse magnetization (right scale) as functions of $q/U_0$ at nonzero temperature. 
	A discontinuous phase transition occurs at $q/U_0 \simeq 0.0256$. 
	The inset shows the thermodynamic potential $J(\bvec{\phi})$, which signifies metastability (see the caption of Fig. \ref{fig:slice} for detail).
	Parameters are $U_1/U_0 = -0.7, zt/U_0 = 0.095, \mu/U_0 = 0.13$, and $k_B T / U_0 = 0.02$. }
	\label{fig:Qdependence_T>0}
\end{figure}

Figures \ref{fig:tdependence_T>0} and \ref{fig:Qdependence_T>0} show the dependences 
of the order parameters $\phi_{\sigma}$ and the transverse magnetization $M_{\rm tr}$
on the hopping amplitude $t$ and the quadratic Zeeman energy $q$. 
It can be seen that we have discontinuous phase transitions between the two superfluid phases. 
In the inset in each figure, where we plot the thermodynamic potential $J(\bvec{\phi})$, 
the presence of multiple local minima indicates the metastability and the accompanying discontinuous phase transition. 
%U_1/U_0 = -0.7
% k_B T / U_0 = 0.02
% \mu/U_0 = 0.13
% The data is taken from C:\Users\sokohaku\Dropbox\doctor_coldatom\SCSposiT\201610251313_u=-0.7,P=0.0,beta=50.0_Qdependence\trial3

Finally, we examine how the temperature affects the nature of phase transitions in this system. 
Figure \ref{fig:boundary_T>0} shows the $t-\mu$ phase diagrams obtained by the binary search algorithm described in the Appendix. 
The temperature is varied between $0$ (upper left panel) and $k_B T /U_0 = 0.05$ (lower right panel). 
For a sufficiently low temperature, the phase diagram is similar to that of the zero-temperature case, 
where we can see two superfluid phases and the normal Bose liquid phase, 
and discontinuous phase transitions occur between the two superfluid phases, when the boundary is close to the normal Bose liquid phase. 
However, by raising the temperature, the regions of discontinuous phase transition shrink and finally disappear, 
i.e., phase transitions between different phases become continuous. 
\begin{figure}
	\centering
	\includegraphics[width = 8cm]{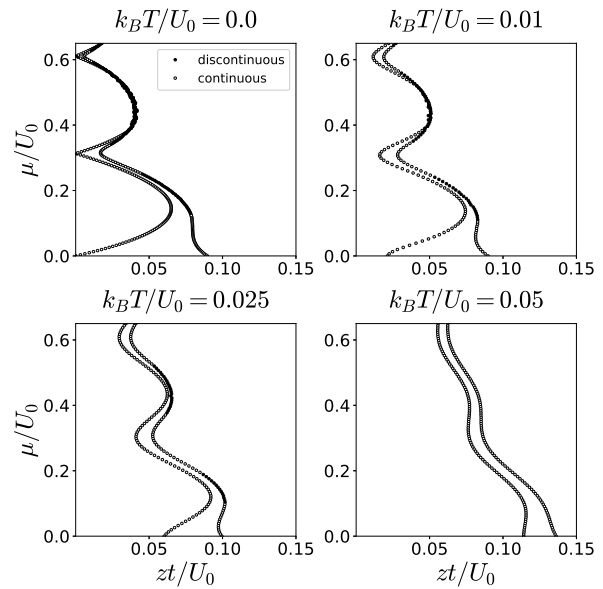}
	\caption{Nonzero-temperature phase diagrams for $(U_1/U_0, q/U_0) = (-0.7, 0.02)$, 
	where the phase boundaries are determined by the binary search algorithm described in the appendix. 
	White circles denote the boundary across which continuous phase transitions occur, while black circles denote the boundary across which discontinuous phase transitions occur. 
	It can be seen that, as the temperature is increased, the regions of discontinuous phase transition shrink, and finally disappear. }
	\label{fig:boundary_T>0}
\end{figure}

%============================================================
\section{Discussions}
\label{sec:disc}

%------------------------------------------------------------
\subsection{On the Mott-insulator phase}
\label{subsec:disc_MI}

Although our calculation shows that the Mott phase has no magnetization and that 
the phase transitions between the Mott insulator phase and the broken-axisymmetry superfluid phase are discontinuous, 
this result should be interpreted with caution. 
This is because the decoupling approximation cannot treat spin fluctuations in the Mott-insulator phase correctly, 
which can be seen from the decoupled grand Hamiltonian (\ref{eq:decoupledK}); in the Mott-insulator phase with $\phi_{\sigma} = 0$, 
the effect of hopping $t$ is completely ignored, 
and it is this hopping term that gives rise to various spin orders in the Mott-insulator phase \cite{ImambekovLukinDemler}. 

A previous study treated the unit-filling Mott-insulator phase by an effective spin model and a field theory \cite{RodriguezEtal}, 
which shows that, inside the Mott lobe, there are two phases, namely, the XY-FM phase and the Ising nematic phase, 
where the rotational symmetry around an applied external magnetic field is broken in the former phase, but not in the latter phase. 
The presence of such a phase transition inside a Mott lobe suggests that 
the discontinuous phase transition from the Mott-insulator phase to the broken-axisymmetry superfluid phase 
predicted in our calculation might be an artifact of the decoupling approximation.

%------------------------------------------------------------
\subsection{Comparison with previous results on spin-1 bosons}
\label{subsec:disc_comparison}

We have found that the ferromagnetic spin-1 bosons in optical lattices can exhibit, in addition to the Mott-insulator phase, two superfluid phases and that 
discontinuous phase transitions occur between these two superfluid phases 
when the hopping amplitude and the quadratic Zeeman energy are varied. 
The presence of two superfluid phases is expected from previous studies on ferromagnetic spin-1 bosons without a lattice 
\cite{KawaguchiUeda,StamperKurnUeda} on the basis of mean-field theory for weak interactions. 
However, the discontinuous phase transition we have found for the lattice system is in clear contrast to the prediction for the system without a lattice; 
in fact, mean-field analyses for the latter system predict continuous phase transitions between these two superfluid phases. 

We note that discontinuous phase transitions between two superfluid phases occur only when the boundary is close to the Mott-insulator phase, 
and that continuous phase transitions occur when the hopping is so large that the boundary is far from the Mott-insulator phase. 
This is consistent with the fact that the corresponding phase transitions are continuous for weakly interacting bosons, 
in which almost all the particles participate in the superfluidity. 
It also suggests that the presence of a considerable portion of the non-superfluid component is crucial for the occurrence of discontinuous phase transitions, 
highlighting the role of the interplay between the spatial degrees of freedom and the spin degrees of freedom.

%------------------------------------------------------------
\subsection{Possible experimental situation}
\label{subsec:disc_exp}

The discontinuous phase transitions between two superfluid phases predicted in our calculation should be experimentally observed 
by examining the presence of hysteresis, as was recently reported for the antiferromagnetic spin-1 bosons in an optical lattice \cite{JiangEtal}. 

First, we note that for the three-dimensional cubic lattice the model parameters can be expressed as 
(see Eqs. (23) and (57) in \cite{KrutitskyPhysRep})
\begin{eqnarray}
	U_0 &=& 4 \sqrt{2\pi} E_R \frac{a_0 + 2 a_2}{3\lambda} x^{3/4},  \\
	U_1 &=& 4 \sqrt{2\pi} E_R \frac{a_2 - a_0}{3\lambda} x^{3/4} , \\
	t &=& \frac{4}{\sqrt{\pi}} E_R x^{3/4} e^{-2\sqrt{x}}, 
\end{eqnarray}
where $x := V_0 / E_R$, 
$a_0$ and $a_2$ are the $s$-wave scattering lengths of the total spin-0 and 2 channels, respectively,  
the optical lattice potential is expressed as $U_L(\bvec{r}) = V_0 \sum_{i=1}^{3} \sin^2 ( \bvec{k}_i \cdot \bvec{r}  )$ 
with $\bvec{k}_i := k \bvec{e}_{i}$ and $\bvec{e}_i$ being the unit vector in the direction $i$, 
$\lambda := 2\pi/k$, 
and $E_R := \hbar^2 k^2 /(2M)$ is the recoil energy with $M$ being the mass of the bosonic atom. 

For ${}^7$Li, we have $a_0 = 39.1 a_B$ and $a_2 = 4.9 a_B$ \cite{VengalattorePrvCom}, where $a_B$ is the Bohr radius, and 
$\lambda = 1064 {\rm nm}$. 
We should take $x \sim 20$ to ensure $zt/U_0 \sim 0.1$. 

Because the magnetization parallel to an applied magnetic field is conserved 
and hence the linear Zeeman effect can be gauged out, 
the quadratic Zeeman energy can easily be tuned with an external magnetic field. 
For example, for ${}^7$Li in the three-dimensional cubic lattice, 
$q/U_0 \simeq 0.1$ corresponds to $B \sim $0.6G. 

As for the temperature, the results in subsection \ref{subsec:results_nzt} indicate that 
the system should be cooled down to as low as $k_B T /U_0 \sim 0.3$, which gives $T \sim$ 3 nK 
for ${}^7$Li in the three-dimensional cubic lattice. 

%============================================================
\section{Conclusion}

We have obtained the zero-temperature and nonzero-temperature phase diagrams of the spin-1 Bose-Hubbard model with the quadratic Zeeman energy. 
Our calculation shows the presence of two distinct superfluid phases, i.e., the broken-axisymmetry superfluid phase and the polar superfluid phase, 
and the discontinuous phase transitions between them for zero temperature and sufficiently low temperature. 

We stress that although the discontinuous superfluid to Mott-insulator phase transitions are known for antiferromagnetic lattice bosons, 
in this paper we found discontinuous phase transitions between superfluid phases, which makes a sharp contrast to the corresponding system without a lattice. 
We hope that our results could be a quantitative guide for further Monte Carlo and experimental studies of ferromagnetic spin-1 bosons.

%============================================================
\section{acknowledgement}

We thank Hosho Katsura for fruitful discussion, and Mukund Vengalattore for valuable private communications. 
This work was supported by 
KAKENHI Grant No. 26287088 from the Japan Society for the Promotion of Science, 
a Grant-in-Aid for Scientific Research on Innovative Areas ``Topological Materials Science'' (KAKENHI Grant No. 15H05855), 
the Photon Frontier Network Program from MEXT of Japan, and the Mitsubishi Foundation. 
K. H. Z. So was supported by the Japan Society for the Promotion of Science through Program for Leading Graduate Schools (ALPS).

\appendix*

%============================================================
\section{Details of the numerical calculations}
\label{sec:app}

%------------------------------------------------------------
\subsection{Matrix elements}
\label{subsec:app_matele}

In performing the numerical calculations, we need matrix elements of $\hat{K}(\bvec{\phi})$ with respect to a specific set of basis. 
As our basis, we take local (i.e. single-site) Fock states $\Phi_{\bvec{n}}$, where $\bvec{n} = (n_1, n _0, n_{-1})$, 
and obtain the following expressions for the matrix elements: 
\begin{eqnarray}
	&{}& \left\langle \Phi_{\bvec{n}},  
		-zt \sum_{\sigma} 
		\left( \phi_\sigma \hconj{a}_{\sigma} + \cconj{\phi_\sigma} a_{\sigma} - |\phi_\sigma|^2 \right) 
		\Phi_{\bvec{n}'} \right\rangle  \nonumber \\
	&=& -zt  \sum_{\sigma} 
			\left( \sqrt{n_\sigma} \phi_\sigma \delta_{\bvec{n}-\bvec{e}_\sigma, \bvec{n}'}
			+ \sqrt{n_\sigma+1} \ \cconj{\phi_\sigma} \delta_{\bvec{n}+\bvec{e}_\sigma, \bvec{n}'}  \right. \nonumber \\
			&{}& \left. - |\phi_\sigma|^2  \delta_{\bvec{n},\bvec{n}' }   \right), 
\end{eqnarray}
\begin{eqnarray}
	&{}& \left\langle \Phi_{\bvec{n}},  
		\left[ \frac{U_0}{2} \hat{n} (\hat{n} - 1) + q (\hat{n}_1 + \hat{n}_{-1}) - \mu \hat{n} \right]
	\Phi_{\bvec{n}'} \right\rangle  \nonumber \\
	&=& \left[ \frac{U_0}{2} n (n - 1) 
				+ q (n_1 + n_{-1})
				- \mu n \right] \delta_{\bvec{n},\bvec{n}' } , 
\end{eqnarray}
\begin{eqnarray}
	&{}& \left\langle \Phi_{\bvec{n}},  
		\frac{U_1}{2} \left( \bvec{S}^{2} - 2 \hat{n} \right) 
	\Phi_{\bvec{n}'} \right\rangle  \nonumber \\
	&=& \frac{U_1}{2} \left\{ \left[ (n_1- n_{-1})^2 + (2n_0-1)(n_1 + n_{-1}) \right]  \delta_{\bvec{n},\bvec{n}'}  \right. \nonumber \\
				&+& 2 \sqrt{n_0(n_0 - 1)(n_1+1)(n_{-1}+1)}  \delta_{\bvec{n}+\bvec{e}_1+\bvec{e}_{-1}-2\bvec{e}_0, \bvec{n}'}  \nonumber \\
				&+& \left. 2 \sqrt{ n_1 n_{-1} (n_0+1)(n_0+2) }  \delta_{\bvec{n}-\bvec{e}_1-\bvec{e}_{-1}+2\bvec{e}_0, \bvec{n}'}  \right\},  
\end{eqnarray}
where $\bvec{e}_{\sigma} := \left( \delta_{\sigma,1}, \delta_{\sigma,0}, \delta_{\sigma,-1} \right)$ 
and $n:= n_1 + n_0 + n_{-1}$.

%------------------------------------------------------------
\subsection{Binary search}
\label{subsec:app_binary}

Here we describe the method used to obtain the results in shown in Figs. \ref{fig:boundary} and \ref{fig:boundary_T>0}. 
Usually, superfluid order parameters grow monotonically with the hopping amplitude $t$. 
(This property is numerically confirmed for the parameter ranges we considered.) 
Then, we can use binary search to locate the phase boundary across which 
the value of $\phi_{\sigma}$ for a specific $\sigma$ changes from 0 to a non-zero value, 
and distinguish discontinuous phase transitions from continuous ones as follows (we denote $zt/U_0$ by $\tau$): 
\begin{enumerate}
	\item Fix the chemical potential .
	\item
		Choose a sufficiently large $\tau_r$, and set $\tau_l = 0$. 
		It is assumed that we have $\phi_\sigma = 0$ at $\tau=\tau_l$ and $\phi_\sigma \neq 0$ at $\tau=\tau_r$ . 
	\item 
		Set $\tau_c := (\tau_r + \tau_l)/2$, solve the self-consistent equation at $\tau=\tau_c$, and obtain $\phi_\sigma$. 
		What we do next depends on whether $\phi_\sigma = 0$ or not
		(Note that because we have inevitable numerical errors in numerical calculations, 
			the condition  $\phi_\sigma \neq 0$ is replaced by $|\phi_\sigma| > \phi_{\rm thresh}$, where we took $\phi_{\rm thresh} \sim 10^{-4}$.): 
		\begin{itemize}
			\item 
				If $\phi_\sigma \neq 0$, the desired value of $\tau$ is equal to or smaller than $\tau_c$, and we substitute $\tau_c$ for $\tau_r$. 
			\item
				If $\phi_\sigma = 0$, the desired value of $\tau$ is equal to or larger than $\tau_c$, and we substitute $\tau_c$ for $\tau_l$. 
		\end{itemize}	
	\item
		Iterate step 3 until $\tau$ converges within a prescribed numerical precision, which we take to be $\sim 10^{-4}$. 
		The final value of $\tau_c$ represents the location of the boundary. 
	\item
		Whether the phase transition is continuous or not can be judged as follows. 
		Define $\phi_r$ and $\phi_l$ as the values of $\phi_{\sigma}$ corresponding to the final values of $\tau_r$ and $\tau_l$, respectively. 
		If the phase transition is continuous, we have $|\phi_r| \simeq |\phi_l|=0$. 
		On the other hand, if the phase transition is discontinuous, $|\phi_r|$ deviates from $|\phi_l|=0$. 
		In our calculation, we take this threshold to be $\sim 10^{-2}$. 
\end{enumerate}

%------------------------------------------------------------
\subsection{Some remarks on the nonzero-temperature case}
\label{subsec:app_notes}

Here, we describe a numerical trick used in the calculations for the nonzero-temperature case. 

Let $U$ be a Hermitian matrix which diagonalizes $\hat{K}(\bvec{\phi})$ as 
$U^{\dag} \hat{K}(\bvec{\phi}) U = {\rm diag}(\lambda_1, \lambda_2, \dots, \lambda_d)$, where $d$ is the dimension of the Hilbert space, 
which is finite because we truncate the Hilbert space by limiting the maximum number of particles per site to $n_{\rm max}$. 
Then, the expectation value of an operator $O$ is 
\begin{equation}
	\langle O \rangle 
	= \frac{  {\rm Tr} [ Oe^{-\beta \hat{K}(\bvec{\phi})} ]  }{ {\rm Tr} e^{-\beta \hat{K}(\bvec{\phi})}  }
	= \frac{ \sum_{i=1}^{d} (U^{\dag} O U)_{i,i} e^{-\beta \lambda_i} }{ \sum_{i=1}^{d} e^{-\beta \lambda_i} }, 
\end{equation}
and the free energy is
\begin{equation}
	J = - \frac{1}{\beta} \log \left( \sum_{i=1}^{d} e^{-\beta \lambda_i} \right). 
\end{equation}

However, these expressions do not work if we implement them directly, because overflow and underflow may occur; 
if all the $\beta\lambda_i$s are large, $\sum_{i=1}^{d} e^{-\beta \lambda_i}$ is regarded as zero, which results in zero-division error, 
and if there is a small (i.e., negatively large) $\beta \lambda_i$, the corresponding $e^{-\beta \lambda_i}$ causes overflow. 
To remedy the situation, we subtract the smallest grand-energy from eigenvalues as follows: 
define $\varepsilon := \min_{i} \lambda_i$, and rewrite
\begin{equation}
	\langle O \rangle 
	= \frac{ \sum_{i=1}^{d} (\hconj{U} O U )_{i,i} e^{-\beta (\lambda_i - \varepsilon)}  }{ \sum_{i=1}^{d} e^{-\beta (\lambda_i - \varepsilon)} }  
\end{equation}
and 
\begin{equation}
	J = \varepsilon - \frac{1}{\beta} \log \left(  \sum_i e^{-\beta (\lambda_i  - \varepsilon) } \right) . 
\end{equation}
By this procedure, all the Boltzmann factors $e^{-\beta \lambda_i}$ are smaller or equal to 1, and at least one $e^{-\beta \lambda_i}$ is 1.

\end{document}